# Laser frequency stabilization and photoacoustic detection based on the tapered fiber coupled crystalline resonator

Yaohui Xu,[1] Xiaolan Liu,[2] Wujun Li,[1] Haotian Wang,[1,3] Jun Guo,[1,4] Jie Ma,[1,5] Jianing Zhang,[1] and Deyuan Shen[1]

[1]*Jiangsu Key Laboratory of Advanced Laser Materials and Devices, School of Physics and Electronic Engineering, Jiangsu Normal University, Xuzhou 221116, China*
[2]*School of Medical Imaging, Xuzhou Medical University, Xuzhou 221116, China*
[3]*wanghaotian@jsnu.edu.cn*
[4]*guojun@jsnu.edu.cn*
[5]*jiema07@jsnu.edu.cn*

**We demonstrate laser frequency stabilization using a high-Q MgF$_2$ crystalline whispering gallery mode resonator coupled with a tapered fiber. We discovered that the tapered fiber, acting as a microcantilever, exhibits mechanical resonance characteristics that is capable of transmitting acoustic perturbations to the frequency locking loop. Both experimental and theoretical investigations into the influence of external acoustic waves on the coupling system were conducted. After acoustic isolation, the locked laser exhibits a minimum frequency noise of 0.4Hz$^2$/Hz at 7kHz and an integral linewidth of 68Hz (0.1s integration time). Benefiting from the ultralow frequency noise of the stabilized laser, it achieves a minimum noise equivalent acoustic signal level of 4.76×10$^{-4}$ Pa/Hz$^{1/2}$. Our results not only facilitate the realization of ultralow noise lasers but also serves as a novel and sensitive photoacoustic detector.**

Ultralow-frequency noise lasers find critical applications in fields such as gravitational wave detection, optical atomic clocks, high-resolution spectroscopy and so on. Typically, laser frequency stabilization employs the Pound-Drever-Hall (PDH) technique to lock the laser frequency to a stable frequency reference, reducing laser power spectral density (PSD) of frequency noise by several orders of magnitude compared to free-running lasers [1]. While atomic and molecular absorption lines are used as compact frequency references, their limited discrete spectral lines make them less suitable for wide-wavelength-range PDH techniques [2]. High-finesse Fabry-Perot cavities [3] and fiber interferometer [4] with long delay lines have been widely employed as frequency references in PDH techniques due to their periodic resonant peaks. However, it typically requires low-temperature and vacuum environments to enhance the stability of these frequency references, resulting in large and complex experimental setups that are mostly found in advanced optical laboratories.

Whispering Gallery Mode Resonators (WGMRs) are characterized by their ultra-high Quality (Q) factors and small volumes. The high Q factors enhance the frequency discrimination accuracy of the reference and monolithic structures offer a natural advantage for miniaturization, particularly in narrow space labs. Specifically, WGMRs made from high-purity crystal materials can achieve Q factors exceeding 10$^{10}$ [5], with the excellent thermal properties of crystals mitigating their intrinsic thermal noise [6]. Therefore, PDH laser frequency stabilization techniques based on crystalline WGMRs have garnered significant interest [6-10]. In most experimental setups utilizing crystalline WGMRs for laser frequency stabilization, prisms are used to couple the free-space laser beam into the cavity. This consideration stems from the good mechanical stability of bulk prisms. To achieve optimal coupling conditions, a narrow air gap is often left between the total internal reflection surface of the prism and the WGMR, creating a suitable evanescent field inside the air gap [11]. This will introduce inherent coupling instability. Direct contact coupling or the use of adhesive to fill the coupling gap, have been proposed to enhance coupling stability but come at the expense of degrading the loaded Q of the WGMRs.

Another common coupling method involves the use of tapered fibers, which conveniently couple the light in the all-fiber configuration, enable the coupling ideality easily [12]. To maximize coupling efficiency, the waist diameter of the tapered fiber is typically reduced to a few micrometers, allowing for sufficient evanescent field to extend into the air cladding. Importantly, when coupling small-sized WGMRs like silica microspheres or microtoroids with tapered fibers, an air gap must also be kept. This is because direct contact of the tapered fiber with the microcavity surface would disrupt the spatial distribution of the WGMs, resulting in a significant reduction in the loaded Q. However, this unfavorable effect applied to crystalline WGMRs with large mode volumes becomes negligible. Furthermore, the electrostatic forces on the WGMR surface naturally stabilize the tapered fiber, ensuring stable contact coupling. The remaining drawback of this approach is that such a light and delicate fiber suspends in the air making it highly sensitive to external disturbances, particularly acoustic waves. This presents a challenge when applying this method to laser frequency stabilization. Nevertheless, we believe that this imperfection in the tapered fiber should not limit its potential to be one of the best solutions for WGMR based laser frequency stabilization.

The motivation of this study is to enhance our understanding of how tapered fibers respond to external acoustic waves, with the aim of advancing the utilization of tapered fiber-coupled WGMR in laser frequency stabilization technology. In practical applications of tapered optical fibers, the taper region is suspended in air to minimize transmission losses caused by contact with other mediums. Thus, we treat the tapered fiber as a microcantilever structure, exhibits strong responses to ambient acoustic waves [13-15]. Moreover, we observed that this microcantilever structure exhibits mechanical resonance characteristics, responding strongly to specific-frequency acoustic waves. External acoustic waves transmitted through the tapered fiber into the WGMR alter the cavity length and coupling strength, ultimately reflected in the PSD of the laser frequency noise. Since the locked laser exhibits extremely low noise in the acoustic frequency range, weak external acoustic signals can also be detected by analyzing the control signals in the laser stabilization system. Therefore, when left open to the environment, our system can also function as a highly sensitive photoacoustic detector.

In this study, we employed a MgF$_2$ crystalline WGMR coupled with a tapered fiber for PDH laser frequency stabilization. To the best of our knowledge, this is the first application of PDH techniques to investigate the response of a tapered fiber coupled with a WGMR to external acoustic waves. Our findings reveal distinct responses in the laser noise PSD for acoustic waves of different frequencies, indicating the mechanical resonance characteristics of the tapered fiber. The coupling system exhibits a minimal noise equivalent acoustic signal level of $4.76 \times 10^{-4}$ Pa/Hz$^{1/2}$, and the locked laser achieves minimal frequency noise of 0.4 Hz$^2$/Hz @7 kHz, with an integral linewidth of 68 Hz. Our results not only develop the WGMR-based laser frequency stabilization technology but also present a novel sensitive photoacoustic detector.

The experimental setup is depicted in Fig. 1(a). The laser from a commercial external cavity diode laser (ECDL, Toptica, CTL 1550) is coupled into an electro-optic modulator (EOM) and then directed to a free-space polarizing controller (PC). This PC comprises a pair of fiber collimators, two 1/4-wave plates, and one 1/2-wave plate. By rotating the waveplates, we can precisely and stably control the polarization state of the output light. Throughout the optical path from the ECDL to the WGMR, polarization-maintaining (PM) fibers are employed to minimize the impact of the external environment on the laser polarization state. The combination of free-space PC and PM fibers is employed to stabilize the incident polarization state into the WGMR, which is crucial for this laser frequency stabilization technique. The laser is further coupled into a WGMR through a tapered fiber. The WGMR used in the experiment is fabricated from precision-polished MgF$_2$ crystal (Evenoptics, HQ-5-M). The loaded Q of the mode used for frequency stabilization is $1.1 \times 10^8$, corresponding to a mode linewidth of ~1.8MHz. With a well-established electronic frequency locking system, the current Q factor is enough for the PDH technology.

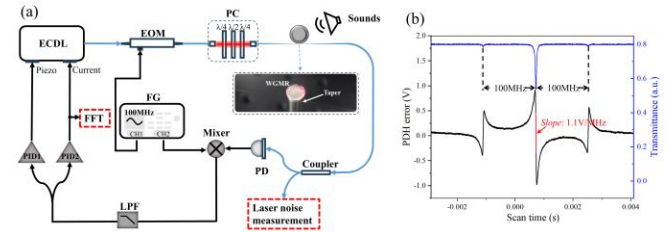

Fig. 1. (a) Experimental setup for laser frequency stabilization and acoustic response monitoring. (b) The PDH error signal and its corresponding WGM transmittance.

The tapered fiber is stably attached to the WGMR surface via electrostatic adhesion (Fig. 1(a) inset). The contact region spans tens of micrometers. Since this region is far smaller than our WGMR mode volume, it only slightly reduces the loaded Q, yet greatly enhances the coupling stability. The laser output from the WGMR is split into two parts using an 80/20 optical coupler. One part is used for laser noise measurement, and the other is connected to a photoelectric detector and then enters the RF port of the electronic mixer. A dual-channel function generator (FG) generates two 100MHz signals, which are used to drive the EOM and the local oscillator (LO) port of the mixer, respectively. The output signal from the IF port of the mixer is passed through a low-pass filter (LPF) and serves as the error signal for laser frequency stabilization. The shape of the error signal depends on the polarization of the light and the phase difference between the EOM driving signal and the RF port of the mixer. Therefore, in the experiment, the error signal is optimized by precisely adjusting the light polarization and controlling a constant and appropriate phase difference between the two channels of the FG. The optimal error signal, as shown in Fig. 1(b), yields a frequency discrimination accuracy of 1.1V/MHz. The error signals are respectively fed into two servo loops, PID1 and PID2. In the slow servo loop (PID1), a piezoelectric ceramic actuator is used, which has a low response bandwidth of ~1kHz. This actuator helps eliminate steady-state errors and laser frequency drift. In the fast servo loop (PID2), a higher bandwidth control feedback is achieved by adjusting the laser diode current, which is used to suppress high-frequency laser noise.

After the laser is locked, a loudspeaker is placed near the WGMR to investigate the influence of external acoustic waves on the coupling system. The intensity and frequency of the loudspeaker can be controlled using an additional signal generator. An acoustic meter is placed equidistant from the loudspeaker to calibrate the sound pressure at the WGMR location. Partial signal from PID2 servo loop is recorded by a 16-bit data acquisition card (NI, USB-6361) then subjected to Fast Fourier Transform (FFT) spectral analysis. Fristly, the loudspeaker generates acoustic waves in the frequency range of 6000-12000Hz with intervals of 500Hz, and the corresponding response spectra are recorded. Subsequently, centered at the frequency of the strongest response, further precise frequency scanning is performed at intervals of 100Hz. In all the experiments above, the sound pressure is maintained at 0.63Pa (90dB). Figs. 2(a) and (b) respectively illustrate the PSDs of the control signals in the PID2 servo under two different tapered fiber coupled to the same WGMR. Clearly, our coupling system exhibits varying responses to acoustic signals of different frequencies. Furthermore, at the same frequency, different tapered fibers exhibit distinct responses, even when they are coupled to the same WGMR. Specifically, the strongest response frequencies of the tapered fibers in Figs. 2(a) and (b) are 9500 and 11200Hz, corresponding to PSD peaks of -22.15 and -23.72dBV/Hz$^{1/2}$, respectively. Therefore, it can be inferred that the acoustic response of our coupling system originates from the tapered fiber rather than the WGMR. The crystalline WGMR is firmly mounted on its pedestal, which makes it less sensitive to acoustic waves when compared to the tapered fiber suspended in air. External acoustic waves can propagate along the tapered fiber into the WGMR, thereby modulating the cavity length and affecting the coupling strength. This effect is ultimately reflected in the PSDs of the control signals. Additionally, the tapered fiber, with its microcantilever-like structure, enhances the intensity of acoustic waves at its natural frequency. The prominent peaks at 9500 and 11200Hz indicate that they correspond to the natural frequency of the tapered fibers in Fig. 2(a) and (b), respectively.

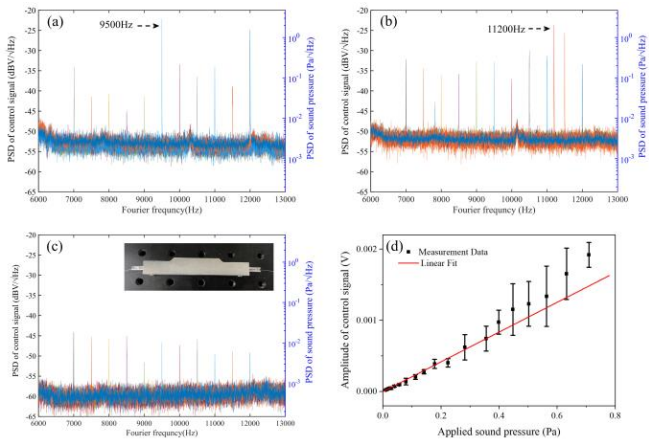

Fig. 2. PSDs of the control signal (PID2) influenced by acoustic waves with different frequency. Different tapered fibers are used in (a) and (b). (c) is the PSD measured after packaging the coupling system and the taped fiber is the same as used in (a). The inset is the picture of the packaged coupling system. (d) Amplitude of the control signal varies with the sound pressure corresponding to 9500Hz acoustic wave in (a).

The tapered fiber can be treated as a resonance problem in a variable cross-section cantilever. Generally, this problem is difficult to have an analytical solution, hence, we employ the finite element numerical method to study this problem. The computational model used is shown in Fig. 3(a), which consists of a tapered region ($2d_2$) and two untapered regions ($d_1$) on both sides. In the experiment, the tapered fiber is glued at both ends, and the coupling region is contact with the WGMR. Considering that the wavelength of acoustic waves is much larger than the fixed region of the tapered fiber, in our computational model, both ends and midpoint of the tapered fiber are considered as fixed points. The introduction of untapered regions in the model is mainly due to practical considerations, where glue is not directly applied to the tapered region to avoid its absorption of light. Visualizations for various mechanical eigenmodes of the tapered fiber are given in Supplement 1. When $d_1$ and $d_2$ are set to 2 and 10.2mm, respectively, the eigenfrequencies $f_0$ of the first three order mechanical modes are 1952, 5015, and 9500 Hz, respectively. We specifically selected the third order mode to calculate the variation of $f_0$ with $d_1$ and $d_2$ due to its closer to the experimental results in Fig. 2(a). The calculation results are shown in Fig. 3(b). As $d_1$ and $d_2$ increase, $f_0$ decreases. In the given parameter range of 1-3mm for $d_1$ and 8-11mm for $d_2$, $f_0$ can vary within the range of 7.5-17.5kHz. The range of $d_1$ and $d_2$ that results in the calculated values of $f_0$ agrees well with our experimental observations. It should be pointed out that $f_0$ is also influenced by the thickness of the tapered fiber. The thinnest waist of the tapered fiber used in the experiment is typically 3-5μm. For convenience, it is set as a fixed value of 3μm in our calculation.

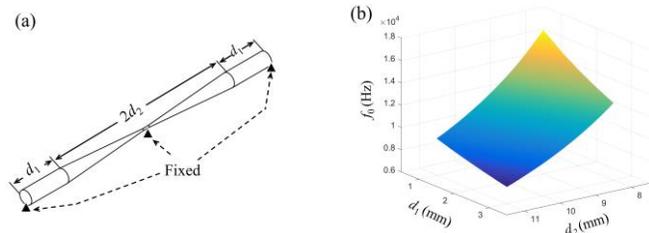

Fig. 3. Schematic diagram of the microcantilever for mechanical mode calculations in the tapered fiber. (b) The resonance frequency of the mechanical modes in the tapered optical fiber varies with the lengths of the untapered ($d_1$) and the tapered regions ($2d_2$).

Next, we examined how the noise peak at 9500Hz in Fig. 2(a) changed in response to variations in the acoustic pressure applied to the coupling system. We conducted five measurements at each sound pressure level and calculated their mean values and standard deviations, as shown in Fig. 2(d). The red line represents the linear fit of the experimental data, yielding a sensitivity of the system to external sound pressure at 0.0021V/Pa. A comparison between the experimental data and the fitted curve shows good agreement at lower sound pressures. However, when the sound pressure exceeded 0.4Pa, the measured data deviated from the fitted line and exhibited a more rapid growth. This phenomenon can be attributed to the resonant effect induced by the 9500Hz acoustic waves on the tapered fiber. It is similar to the threshold effect observed in laser systems, where the light-light curve experiences a significant increase in slope when the pump power exceeds the laser threshold. Additionally, based on the acoustic sensitivity obtained from Fig. 2(d), it is possible to convert the PSD of the control signal into that of sound pressure, as indicated on the right $y$-axis of Fig. 2. Consequently, by utilizing the noise floor of the control signal PSD (∼-53dBV/Hz$^{1/2}$), the Noise Equivalent Acoustic Signal Level (NEASL) is calculated to be 0.004Pa/Hz$^{1/2}$, which is attributed to the noise of the surroundings.

To isolate surrounding acoustic waves, we carefully package the coupling system. An aluminum box whose inner wall is attached by soundproofing felt encloses the WGMR and tapered fiber, any gaps are fully sealed. Inside this aluminum enclosure, the tapered fiber is secured at multiple fixed points along its axis to dissipate external mechanical vibrations transmitted through the input and output fiber pigtails. Unlike conventional vacuum chamber enclosures, this design is selected to accommodate various experimental environments. To exploit the advantage of the small size of the WGMR, our design also aims to minimize the packaging volume. The packaged system has dimensions of 100mm×15mm×15mm and is compatible with other fiber components, as illustrated in the inset of Fig. 2(c). The measured control signal PSD with respect to external acoustic waves after package is given in Fig. 2(c). Consequently, even under the same sound pressure used in Figs. 2(a) and (b), the acoustic noise peaks for all frequencies are controlled below -45dBV/Hz$^{1/2}$. The noise floor of the PSD of the is -60dBV/Hz$^{1/2}$, which is 7dB lower compared to the unpackage system. Accordingly, the calculated NEASL is $4.76\times10^{-4}$Pa/Hz$^{1/2}$ for the packaged system. This implies that our packaged system is capable of isolating ambient noise effectively.

Finally, we employed the packaged tapered fiber and WGMR to achieve laser frequency stabilization. We also employ a thermoelectric cooler (TEC) to ensure precise temperature control of the aluminum box. Fig.4 presents the PSD of frequency noises for the laser before and after locking, respectively. As indicated by the red curve, the locked laser demonstrates a minimum frequency noise of 0.4Hz$^2$/Hz at 7kHz. The frequency noise of the free-running laser is suppressed by more than 50 dB over an offset frequency range from 10$^1$ to 10$^3$ Hz. The integrated linewidth of the locked laser is estimated via the β-line method to be 68Hz (0.1s integration time). The presence of numerous peaks in the frequency of 10$^2$-10$^3$Hz is attributed to residual noises from the vibrations on the table as well as the interferometer used for frequency noise measurements. The noise below 10$^2$Hz is caused by the fundamental thermal noise inside the MgF$_2$ crystalline WGMR itself [16]. The rapid increase in frequency noise beyond 10$^4$Hz is attributed to the electronic noise from the PDH frequency locking circuits.

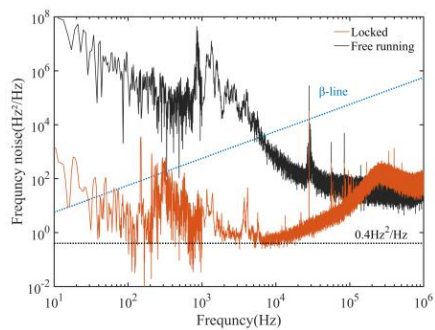

Fig. 4. Laser frequency noise before and after locking.

In conclusion, we have demonstrated the resonant response of the tapered fiber to acoustic waves in the WGMR-based PDH laser locking system. We experimentally observed that acoustic waves influence the PSD of the frequency noise through the tapered fiber, exhibiting clear characteristics of mechanical resonance. With the soundproofed packaging of the tapered fiber and the WGMR, we achieved a minimum laser noise of 0.4Hz$^2$/Hz and an integral linewidth of 68Hz after locking. Moreover, this system serves as a novel and sensitive photoacoustic detector, being capable of a noise equivalent acoustic signal level of $4.76\times10^{-4}$Pa/Hz$^{1/2}$. Investigating to the acoustic response of tapered fibers in laser frequency stabilization systems holds significant promise for the future development of miniaturized WGMR-based PDH techniques.

**Funding.** National Natural Science Foundation of China (61805112, 62035007, 62075089), National Key Research and Development Program of China (2022YFB3605800), Natural Science Foundation of Xuzhou (KC22296).

**Disclosures.** The authors declare no conflicts of interest.

**Supplemental document.** See Supplement 1 for supporting content.

### References

1. E. D. Black, "An introduction to Pound–Drever–Hall laser frequency stabilization," *Am. J. Phys.* **69**, 79 (2001).
2. C. Tamm, S. Weyers, B. Lipphardt, *et al.*, "Stray-field-induced quadrupole shift and absolute frequency of the 688-THz 171Yb+ single-ion optical frequency standard," *Phys. Rev. A.* **80**, 043403 (2009).
3. T. Kessler, C. Hagemann, C. Grebing, *et al.*, "A sub-40-mHz-linewidth laser based on a silicon single-crystal optical cavity," *Nat. Photonics.* **6**, 687-692 (2012).
4. F. Kéfélian, H. F. Jiang, P. Lemonde, *et al.*, "Ultralow-frequency-noise stabilization of a laser by locking to an optical fiber-delay line," *Opt. Lett.* **34**, 914-916 (2009).
5. I. S. Grudinin, V. S. Ilchenko, and L. Maleki, "Ultrahigh optical Q factors of crystalline resonators in the linear regime," *Phys. Rev. A* **74**, 063806 (2006).
6. J. Alnis, A. Schliesser, C. Y. Wang, *et al.*, "Thermal-noise-limited crystalline whispering-gallery-mode resonator for laser stabilization," *Phys. Rev. A.* **84**, 011804 (2011).
7. I. Fescenko, J. Alnis, A. Schliesser, *et al.*, "Dual-mode temperature compensation technique for laser stabilization to a crystalline whispering gallery mode resonator," *Opt. Express.* **20**, 19185-19193 (2012).
8. J. K. Lim, A. A. Savchenkov, E. Dale, *et al.*, "Chasing the thermodynamical noise limit in whispering-gallery-mode resonators for ultrastable laser frequency stabilization," *Nat Commun.* **8**, 8 (2017).
9. J. K. Lim, A. A. Savchenkov, A. B. Matsko, *et al.*, "Microresonator-stabilized extended-cavity diode laser for supercavity frequency stabilization," *Opt. Lett.* **42**, 1429-1452 (2017).
10. M. S. D. Cumis, S. Borri, G. Insero, et al., "Microcavity-Stabilized Quantum Cascade Laser," *Laser Photonics Rev.* **10**, No. 1, 153–157 (2016).
11. M. L. Gorodetsky and V. S. Ilchenko, "Optical microsphere resonators: optimal coupling to high-Q whispering-gallery modes," *J. Opt. Soc. Am. B* **16**, 147 (1999).
12. S. M. Spillane, T. J. Kippenberg, O. J Painter, *et al.*, "Ideality in a Fiber-Taper-Coupled Microresonator System for Application to Cavity Quantum Electrodynamics," *Phys. Rev. Lett.* **91**, 043902 (2003).
13. B. Xu, Y. Li, M. Sun, *et al.* "Acoustic vibration sensor based on nonadiabatic tapered fibers," *Opt. Lett.* **37**, 4768 (2012).
14. L. Su, and S. R. Elliott, "All-fiber microcantilever sensor monitored by a low-cost fiber-to-tip structure with subnanometer resolution," *Opt. Lett.* **35**, 1212 (2010).
15. Y. Li, X. Wang, and X. Bao, "Sensitive acoustic vibration sensor using single-mode fiber tapers," *Appl. Opt.* **50**, 1873 (2011).
16. A. B. Matsko, A. A. Savchenkov, N. Yu, et al., "Whispering-gallery-mode resonators as frequency references. I. Fundamental limitations," J. Opt. Soc. Am. B. **24**, 1324-1335 (2007).